\documentclass[aps,prd,twocolumn,superscriptaddress,showkeys,nofootinbib]{revtex4-2}

\usepackage[utf8]{inputenc}
\usepackage{amsmath}
\usepackage{amssymb}
\usepackage[mathscr]{eucal}
\usepackage{graphicx}
\usepackage{xcolor}

\usepackage{color}
\usepackage{colortbl}
\usepackage{multirow}
\definecolor{royes}{HTML}{ff0000}
\definecolor{rop}{HTML}{ffff00}
\definecolor{rono}{HTML}{7cfc00}

\def\fs#1{\setbox0=\hbox{$#1$}#1\hskip-\wd0\dimen0=5pt\advance\dimen0
by-\ht0\advance\dimen0 by\dp0\lower0.5\dimen0\hbox to\wd0{\hss\sl/\/\hss}}

\begin{document}

\title{\LARGE Proton decay}

\author{Tommy Ohlsson}
\email[]{tohlsson@kth.se}
\affiliation{Department of Physics, School of Engineering Sciences, KTH Royal Institute of Technology, Roslagstullsbacken 21, SE--106~91 Stockholm, Sweden}
\affiliation{The Oskar Klein Centre for Cosmoparticle Physics, AlbaNova University Center, Roslagstullsbacken 21, SE--106~91 Stockholm, Sweden}

\begin{abstract}
Proton decay is a hypothetical form of particle decay in which protons are assumed to decay into lighter particles. This form of decay has yet to be detected. In this contribution to the proceedings of Neutrino 2022, we review the current status of proton decay, covering both experimental results and theoretical models, including their predictions.
\end{abstract}

\maketitle

\section{Introduction}
\label{sec:intro}

We review the present status of hypothetic proton decay. We discuss both past and future experimental efforts as well as theoretical development. Especially, we consider proton decay in so-called grand unified theories.

This work is organized as follows. First, in Sec.~\ref{sec:wipd?}, we will address the question: What is proton decay? In Sec.~\ref{sec:eod}, an estimate of decay in general will be given. Then, in Sec.~\ref{sec:hopd}, the history of proton decay, including results by experiments, will be discussed. Next, in Sec.~\ref{sec:pdit}, proton decay in theory (in so-called basic GUTs) will be described and estimates of proton lifetime will be presented. In Sec.~\ref{sec:pdinsg}, we will shortly study proton decay in non-SUSY GUTs [{\it e.g.} SU(5) and SO(10)]. In addition, in Sec.~\ref{sec:2017now}, we will review SU(5) and SO(10) models that have been developed in the literature in the last five years. In Sec.~\ref{sec:fe}, the most promising future experiments will be mentioned. Finally, in Sec.~\ref{sec:s&o}, a personal summary and outlook will be presented.

\section{What is proton decay?}
\label{sec:wipd?}

In particle physics, proton decay is a {\em hypothetical form} of particle decay in which protons decay into lighter subatomic particles. Examples of potential proton decay channels are:
\begin{itemize}
\item $p \to e^+ + \pi^0$, \quad $p \to \mu^+ + \pi^0$ \quad (canonical examples)
\item $p \to \overline{\nu} + \pi^+$, \quad $p \to \mu^+ + K^0$, \quad $p \to \overline{\nu} + K^+$, \quad \ldots
\end{itemize}
On the other hand, positron emission (or $\beta^+$ decay) $p \to n + e^+ + \nu_e$ and electron capture $p + e^- \to n + \nu_e$ are {\em not} examples of proton decay, since the protons in these processes interact with other subatomic particles inside nuclei or atoms.

\section{Estimate of decay}
\label{sec:eod}

\begin{figure}
\includegraphics[width=\columnwidth]{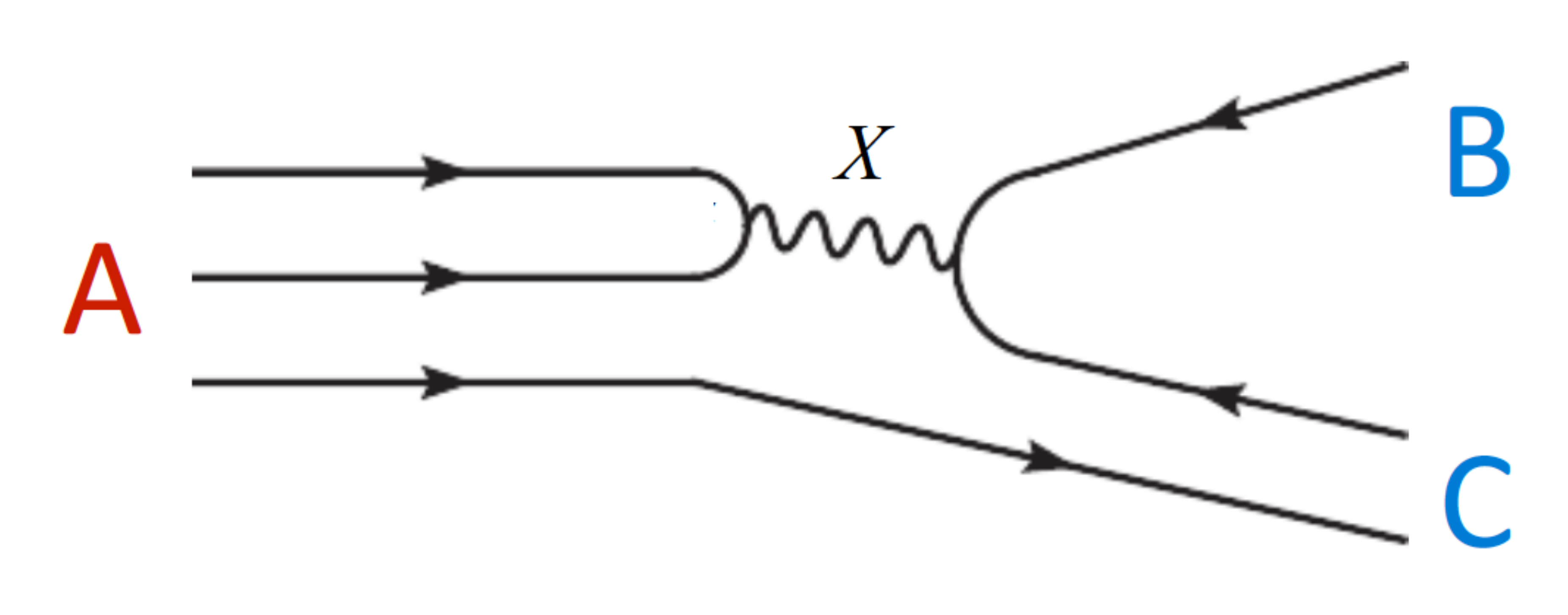}
\caption{Decay of ${\color{red} A}$ into ${\color{blue} B}$ and ${\color{blue} C}$ through exchange of $X$}
\label{fig:decay}
\end{figure}

First, we discuss how to calculate the lifetime of a decaying particle $A$ in the decay ${\color{red} A} \to {\color{blue} B} + {\color{blue} C}$ in theory, see Fig.~\ref{fig:decay}. 
Assuming $M_X \gg m_A$, where $M_X$ and $m_A$ are the masses of an exchange particle $X$ and the initial decaying particle $A$, respectively, the Feynman amplitude for the decay can be approximated by an effective four-fermion interaction, namely
\begin{equation}
\langle {\color{blue} BC} | {\color{red} A} \rangle \simeq g \cdot \frac{1}{M_X^2} \cdot g = \frac{g^2}{M_X^2} \propto \frac{\alpha}{M_X^2}, \qquad \alpha \equiv \frac{g^2}{4\pi},
\end{equation}
where $g$ is the coupling constant of the interaction. Now, a rough estimate of the total decay width is
\begin{equation}
\Gamma({\color{red} A} \to {\color{blue} BC}) \simeq |\langle {\color{blue} BC} | {\color{red} A} \rangle|^2 m_A^5 \propto \frac{\alpha^2}{M_X^4} m_A^5.
\end{equation}
Thus, on dimensional grounds, the lifetime of the decaying particle $A$ is given by
\begin{equation}
\tau_A \equiv \frac{1}{\Gamma({\color{red} A} \to {\color{blue} BC})} \propto \frac{M_X^4}{\alpha^2 m_A^5}.
\label{eq:tauA}
\end{equation}

\section{History of proton decay}
\label{sec:hopd}

Second, we discuss how to measure proton decay in experiments. For example, the potential decay channel $p \to e^+ + \pi^0$ could be searched for in the following process:
\begin{align}
p \longrightarrow e^+ + \; & \pi^0 \nonumber\\
	& \hookrightarrow \gamma + \gamma. \nonumber
\end{align}
\begin{figure}
\includegraphics[clip,trim=0 6.5cm 0 6.5cm,width=4cm,width=0.75\columnwidth]{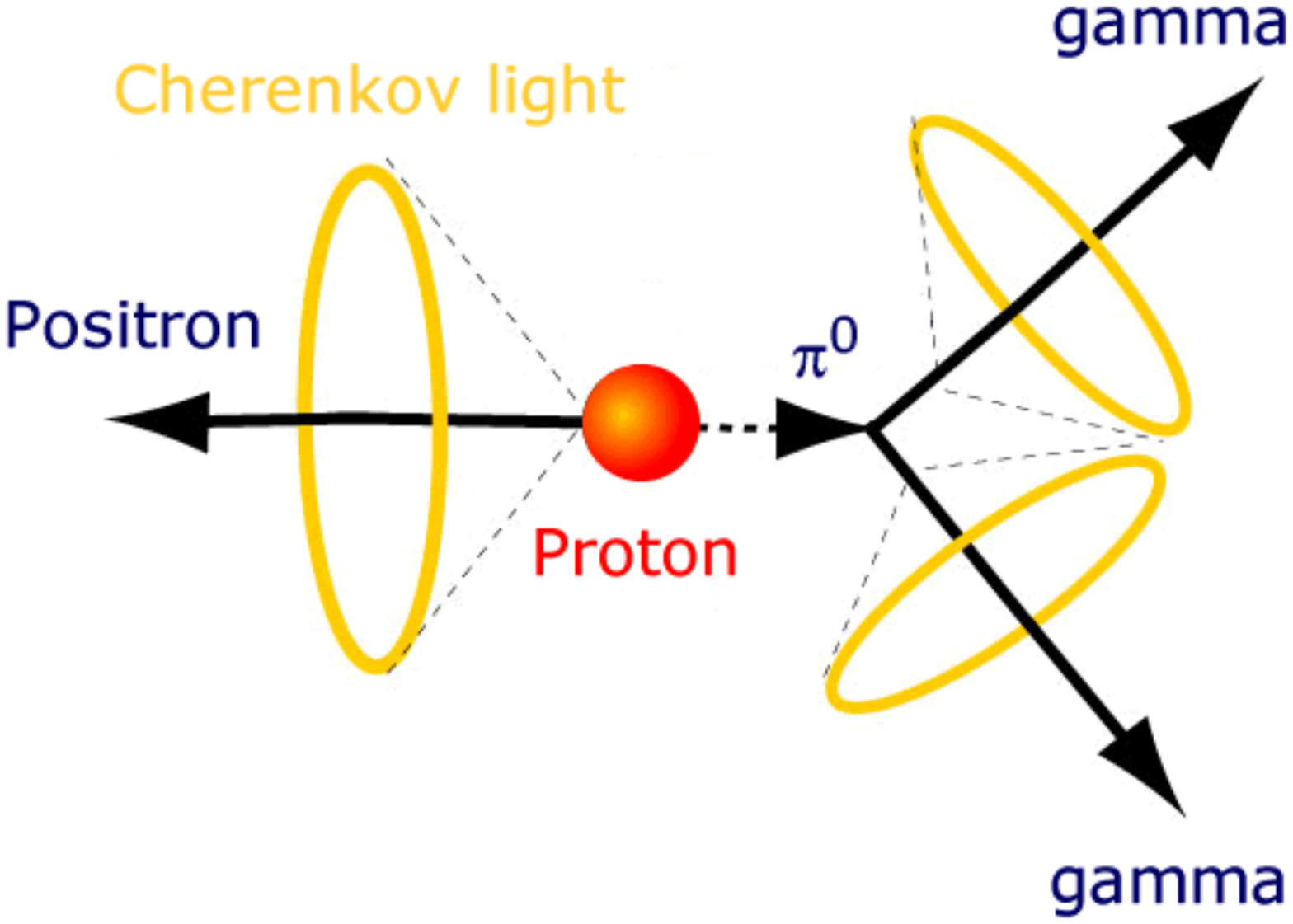}
\caption{Proton decay in the potential decay channel $p \to e^+ + \pi^0$ with Cherenkov light}
\label{fig:proton_decay_cherenkov}
\end{figure}
In a water-Cherenkov detector, the two photons would be detected as two rings of Cherenkov light and the positron would also produce a third ring by Cherenkov radiation, see Fig.~\ref{fig:proton_decay_cherenkov}. Thus, the signal would be measured through {\em three rings} of Cherenkov light.

\begin{figure}
\includegraphics[width=\columnwidth]{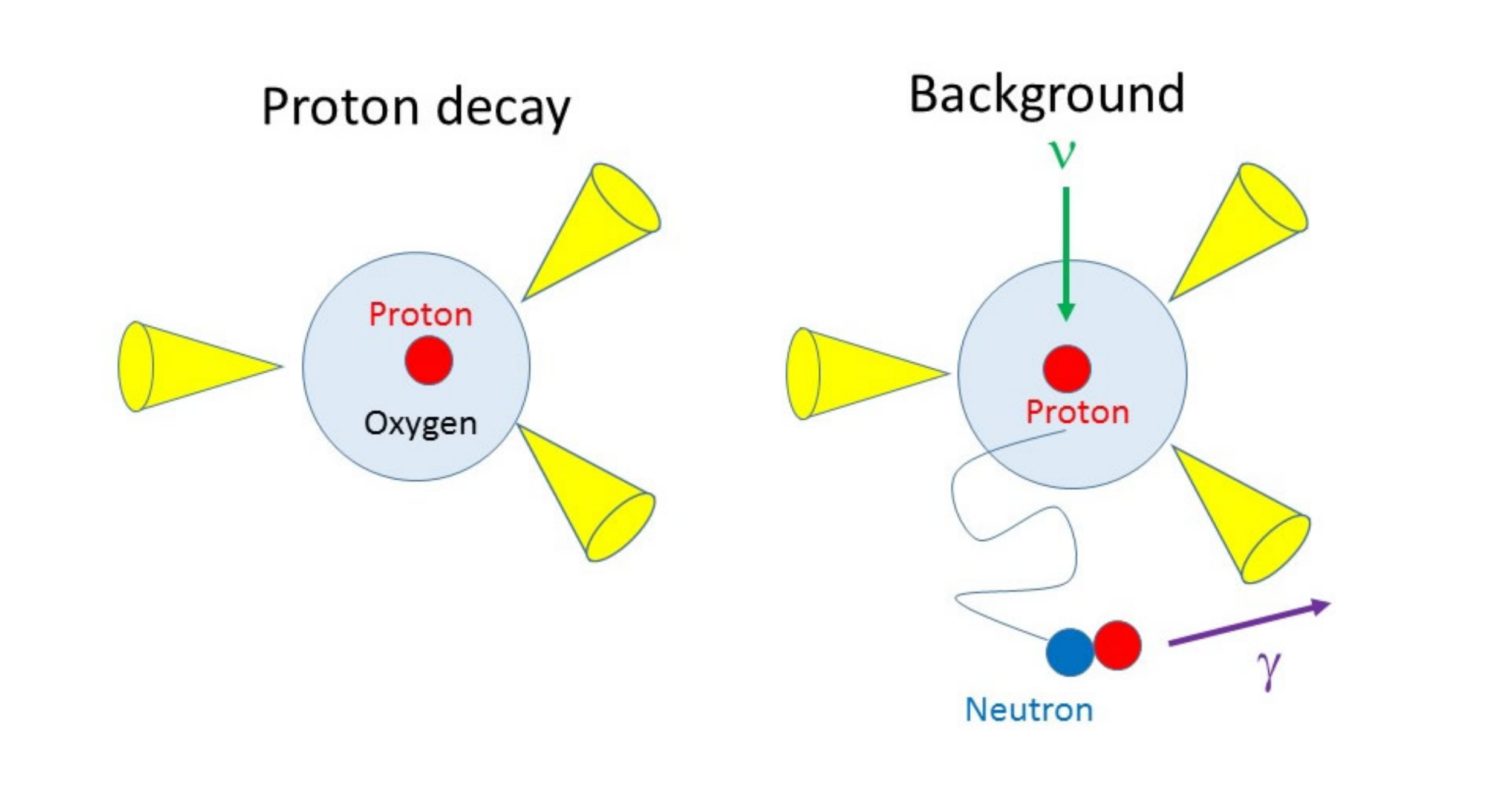}
\caption{Proton decay and background: The left sketch shows a signal event (`Proton decay'), whereas the right sketch shows a background event (`Background').}
\label{fig:proton_decay_background}
\end{figure}

The major background to the signal of proton decay consists of events created by atmospheric neutrinos. How does an event look like in the proton decay channel $p \to e^+ + \pi^0$? As discussed above, the signal is given by the process $p \to e^+ + \pi^0 \to e^+ + \gamma + \gamma$, which produces three rings of Cherenkov radiation (see the left sketch in Fig.~\ref{fig:proton_decay_background}), where one ring comes from $e^+$ and two rings come from $\pi^0 \to \gamma + \gamma$. On the other hand, the major background is created by the process $\nu_e + p \to e^- + n + \pi^0 \to e^- + n + \gamma + \gamma$, where $\nu_e$ stems from atmospheric neutrinos that are produced through the processes $\pi^+ \to \mu^+ + \nu_\mu$ and $\mu^+ \to e^+ + \overline{\nu_\mu} + \nu_e$ when cosmic rays hit the Earth's atmosphere.
Thus, atmospheric neutrino events \underline{mimic} ``proton decay'' events. In a background event (see the right sketch in Fig.~\ref{fig:proton_decay_background}), a neutron $n$ does not produce a ring, but it is sometimes captured by a proton $p$ followed by delayed gamma ray emission $\gamma$. Therefore, the major background events are characterized by three rings of Cherenkov radiation and potential gamma ray emission.

In the past, there are some experiments that have been searching for proton decay. These experiments divide into (i) water-Cherenkov detectors such as {\sc IMB} (Ohio, USA; 1982--1991) and {\sc KamiokaNDE} (Kamioka Nucleon Decay Experiment, Gifu, Japan; 1983--1985, 1985--1990, 1990--1995) as well as (ii) iron-tracking calorimeters such as {\sc NUSEX} (Nucleon Stability Experiment, Mont-Blanc, France; 1982--1983), which found a candidate event in the $p \to \mu^+ + K^0$ channel, {\sc Fr{\'e}jus} (Fr{\'e}jus, France; 1984--1988), and {\sc Soudan} (Minnesota, USA; 1981--1982, 1989--2001), which also found a candidate event in the $p \to \overline{\nu} + K^+$ channel. Presently, there is one operating experiment, which is {\sc Super-Kamiokande} (water-Cherenkov detector, Gifu, Japan; 1996--now). In general, concerning the water-Cherenkov detectors, the experiments with respect to the search of proton decay can be summarized as follows:
\begin{itemize}
\item {\sc IMB}: 3.3~kton (fid.~vol.), 2~000 PMTs (4~\%).\\ No proton decay have been found,
$\tau(p \to e^+ \pi^0) > 5.5 \cdot 10^{32}$~years \cite{Gajewski:1989gh}.
\item {\sc KamiokaNDE}: 0.88~kton (fid.~vol.), 948 PMTs (20~\%).\\ No proton decay have been found,
$\tau(p \to e^+ \pi^0) > 2.6 \cdot 10^{32}$~years \cite{Kamiokande-II:1989avz}.
\item {\sc Super-Kamiokande}: 22.5~kton (fid.~vol.), 11~146 PMTs (40~\%).\\ No proton decay have been found, still operating.
\end{itemize}
In particular up to now, the results on proton decay by {\sc Super-Kamiokande} (SK) \cite{Super-Kamiokande:2020wjk} can be summarized as follows. First, in the $p \to e^+ + \pi^0$ channel, {\em no} candidate events have been found, which leads to the current and best lower bound on the proton lifetime, {\it i.e.}
$$
\tau(p \to e^+ \pi^0) > 2.4 \cdot 10^{34} \, \mbox{years} \quad @~90~\%~{\rm C.L.}
$$
Second, in the $p \to \mu^+ + \pi^0$ channel, {\em one} candidate event remains in data, which despite that means that a lower bound on the proton lifetime has been possible to obtain, which is given by
$$
\tau(p \to \mu^+ \pi^0) > 1.6 \cdot 10^{34} \, \mbox{years} \quad @~90~\%~{\rm C.L.}
$$
Finally, in Tab.~\ref{tab:SK}, earlier results by SK on lower bounds on proton lifetime in different potential decay channels are presented.
\begin{table*}
\begin{tabular}{l r r}
\hline
\hline
\multicolumn{2}{c}{Lower bounds on proton lifetime [years]} & Reference\\
\hline
$\tau(p \to 3\ell) > 0.92 \cdot 10^{34}$ & & Ref.~\cite{Super-Kamiokande:2020tor}\\
$\tau(p \to e^+ \pi^0) > 1.6 \cdot 10^{34}$ & $\tau(p \to \mu^+ \pi^0) > 7.7 \cdot 10^{33}$ & Ref.~\cite{Super-Kamiokande:2016exg}\\
$\tau(p \to \overline{\nu} K^+) > 5.9 \cdot 10^{33}$ & & Ref.~\cite{Super-Kamiokande:2014otb}\\
$\tau(p \to \mu^+ K^0) > 1.6 \cdot 10^{33}$ & Poster by R.~Matsumoto:~$> 3.6 \cdot 10^{33}$ & Ref.~\cite{Super-Kamiokande:2012zik}\\
$\tau(p \to e^+ \pi^0) > 8.2 \cdot 10^{33}$ & $\tau(p \to \mu^+ \pi^0) > 6.6 \cdot 10^{33}$ & Ref.~\cite{Super-Kamiokande:2009yit}\\
$\tau(p \to \overline{\nu} K^+) > 6.7 \cdot 10^{32}$ & & Ref.~\cite{Super-Kamiokande:1999xld}\\
$\tau(p \to e^+ \pi^0) > 1.6 \cdot 10^{33}$ & & Ref.~\cite{Super-Kamiokande:1998mae}\\
\hline
\hline
\end{tabular}
\caption{Earlier results by SK on lower bounds on proton lifetime $@~90~\%~{\rm C.L.}$ in different potential decay channels.}
\label{tab:SK}
\end{table*}

\section{Proton decay in theory}
\label{sec:pdit}

In the Standard Model (SM), baryon number $B$ and lepton number $L$ are conserved.\footnote{Baryon and lepton numbers are ``conserved'' in the Standard Model, but they can, in fact, be violated by chiral anomalies, since there are problems to apply these symmetries universally over all energy scales. In any case, note that their difference $B-L$ is conserved.} For example, the decay channel $p \to e^+ + \pi^0$ is \underline{forbidden} in the SM, since both $B$ and $L$ are {\bf not} conserved in this process:
\begin{center}
\begin{tabular}{r c c c c c}
 & $p$ & $\longrightarrow$ & $e^+$ & $+$ & $\pi^0$ \\[1mm]
$B:$ & $1$ & $\neq$ & $0$ & $+$ & $0$ \\
$L:$ & $0$ & $\neq$ & $-1$ & $+$ & $0$ \\
$B - L:$ & $1$ & $=$ & $1$ & $+$ & $0$ \\
\end{tabular}
\end{center}
In fact, the proton $p$ is \underline{stable} in the SM, {\it i.e.}
\begin{equation}
\Gamma(p \to \cdots) = 0 \quad \Longrightarrow \quad \tau_p \to \infty.
\end{equation}
In the SM (with minimal particle content), $B$ and $L$ are conserved due to {\it gauge invariance} and {\it renormalizability} which ensure that $B$ and $L$ are global symmetries of the theory \cite{Nanopoulos:1973wz,Weinberg:1973un}. In the SM at the non-renormalizable level, there are at least {\bf two} obvious possibilities for \underline{violation} of $B$ and $L$:
\begin{enumerate}
\item The particle content of the SM is enlarged.
\item The gauge group ${\rm SU(3) \times SU(2) \times U(1)}$ of the SM is extended.
\end{enumerate}
This leads to effective operators that can cause violation of $B$ and $L$ which are non-renormalizable and have dimensions $d > 4$ as well as coupling constants with dimensions $[M^{4-d}]$. In the SM at the non-renormalizable level, effective operators $d > 4$ can be summarized as the following three dimensional categories:
\begin{itemize}
\item \underline{Dimension $d = 5$:} The {\it Weinberg operator} is the only operator, which can be formally written as \cite{Weinberg:1979sa}
\begin{equation}
{\mathscr L}_5 \sim \frac{1}{M} (H \ell) (H \ell).
\end{equation}
This operator has $\Delta B = 0$ and $\Delta L = 2$ and gives rise to Majorana neutrino masses.
See talk on {\it Theoretical models of neutrino masses} by F.~Feruglio.
\item \underline{Dimension $d = 6$:} Four-fermion operators can be schematically written as \cite{Weinberg:1979sa,Wilczek:1979hc,Abbott:1980zj}
\begin{equation}
{\mathscr L}_6 \sim \frac{1}{M^2} qqq\ell + {\rm h.c.}
\end{equation}
These operators have $\Delta B = 1$ and $\Delta L = 1$, but $\Delta (B-L) = 0$, which means that they violate $B+L$ but conserve $B-L$.
All dimension-six operators lead to two-body proton decay.
\item \underline{Dimension $d \geq 7$:} These operators are naturally suppressed with respect to dimension-six operators.
\end{itemize}
In 1967, the {\em hypothesis of proton decay} was formulated by Sakharov. Three necessary conditions for baryon asymmetry ({\it i.e.}~generation of a non-zero baryon number in the initially matter-antimatter symmetric Universe) were proposed \cite{Sakharov:1967dj}. The three {\it Sakharov conditions} are:
\begin{enumerate}
\item Baryon number violation $\Delta B \neq 0$
\item $C$- and $CP$-violation
\item Interactions out of thermal equilibrium
\end{enumerate}

In fact, only physics beyond the SM can connect higher-dimensional effective operators with other physical phenomena and thus be testable. The best candidate for physics beyond the SM that involves proton decay is grand unified theories (GUTs). Other candidates are quantum tunneling, quantum gravity, extra dimensions, string theory, {\it etc.} Now, the main problem is that there is a plethora of many different GUTs.

In general, GUTs predict exotic interactions through their additional gauge bosons and scalars, which can mediate proton decay. The gauge boson-mediated proton decay can be observed in the covariant derivative of the fermions in which gauge bosons couple to both quarks and leptons. Therefore, they are called {\it leptoquark} gauge bosons and can convert quarks to leptons, and vice versa. The leptoquark gauge and scalar bosons violate $B$. These leptoquark gauge bosons can be integrated out, and thus, {\em effective dimension-six operators} are produced that describe proton decay. An inherent problem is model dependence of GUTs. First, proton decay is a generic prediction of GUTs, but there is some {\em model dependence} in the allowed effective operators, depending on which couplings are present in the given GUT. Second, there is a difference in GUTs with or without supersymmetry (SUSY). In addition to dimension-six operators, SUSY allows for {\em dimension-five and dimension-four operators} that may lead to faster proton decay.

To sum up, proton decay operators in GUTs are divided into the following three types of dimensional operators:
\begin{itemize}
\item \underline{Dimension-six operators:} All of the dimension-six proton decay operators violate {\bf both} $B$ and $L$ but not $B - L$. The exchange of leptoquark gauge bosons (denoted $X$ or $Y$) with masses $M_{\rm GUT}$ can lead to that some operators are suppressed by $1/M_{\rm GUT}^2$. The exchange of triplet Higgs with mass $m_H$ can lead to that all of the operators are suppressed by $1/m_H^2$.
\item \underline{Dimension-five operators:} In SUSY GUTs, it is also possible to have dimension-five proton decay operators. In this case, the operators are suppressed by $(M M_{\rm SUSY})^{-1}$, where $M$ is the mass of the exchange particle and $M_{\rm SUSY}$ is the mass scale of the superpartners.
\item \underline{Dimension-four operators:} In SUSY GUTs with the absence of $R$-parity, it is also possible to have dimension-four proton decay operators. In this case, the operators are suppressed by $1/M_{\rm SUSY}^2$ which gives rise to proton decay that is normally too fast.
\end{itemize}
In GUTs, since all dimension-six proton decay operators conserve $B-L$, a {\it proton} {\bf always} decays into an {\it antilepton}.
\begin{figure}
\includegraphics[width=\columnwidth]{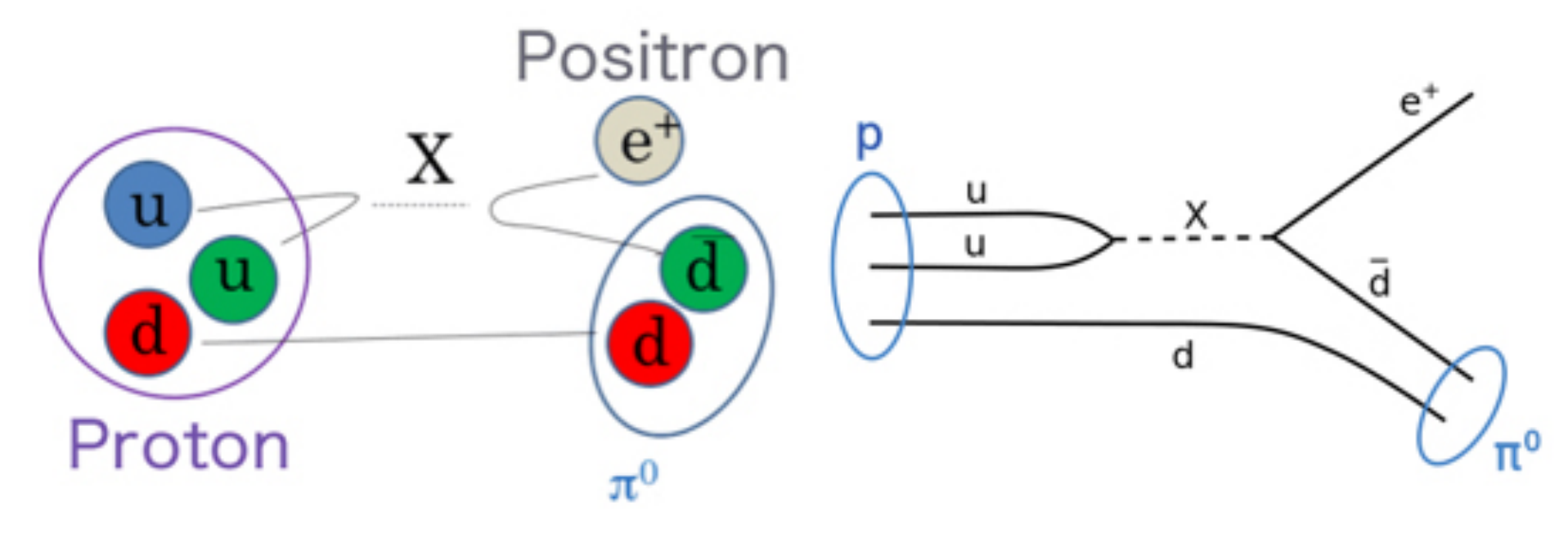}
\caption{The potential proton decay channel $p \to e^+ + \pi^0$}
\label{fig:pdecay_X}
\end{figure}
As already discussed, a generic example of proton decay in GUTs is given by the potential decay channel $p \longrightarrow e^+ + \pi^0$ (see Fig.~\ref{fig:pdecay_X}), and using Eq.~\eqref{eq:tauA}, a rough estimate of the proton lifetime with a GUT-scale gauge boson mediator (like $X$) is given by
\begin{equation}
\tau_p \propto \frac{M_{\rm GUT}^4}{\alpha_{\rm GUT}^2 m_p^5},
\label{eq:taup}
\end{equation}
where $m_p$ is the mass of the proton and $\alpha_{\rm GUT}$ is the gauge coupling constant at the GUT-scale $M_{\rm GUT}$. Now, a long proton lifetime requires a combination of a large mass scale $M_{\rm GUT}$ and a small coupling constant $\alpha_{\rm GUT}$, since $\tau_p \propto M_{\rm GUT}^4/\alpha_{\rm GUT}^2$. From Eq.~\eqref{eq:taup}, we observe that there is a stronger dependence on $M_{\rm GUT}$ than $\alpha_{\rm GUT}$. Therefore, non-observation of proton decay leads to a lower bound on $M_{\rm GUT}$.

It is possible to derive a more precise estimate of the proton lifetime in GUTs. In order to perform such a derivation, we make three assumptions. First, we assume that the proton contains physical quark mass states, which means that a Yukawa mixing factor should be included. Second, the proton is given at the scale $M \simeq 1$~GeV, whereas operators are at scale $M \simeq M_{\rm GUT}$, which means that a renormalization factor due to renormalization group running between $1$~GeV and $M_{\rm GUT}$ should be included. Third, projection of the proton state onto the meson state means that a hadronic matrix element should be included. Using the three assumptions, we find the proton decay width for the channel $p \to e^+ + \pi^0$ as \cite{Nath:2006ut}
\begin{equation}
\Gamma(p \to e^+ \pi^0) \simeq \frac{\pi m_p}{4 f_\pi^2} \frac{\alpha_{\rm GUT}^2}{M_{\rm GUT}^4} F_q R^2 \alpha_H^2,
\end{equation}
where $f_\pi$ is the pion decay constant, $F_q$ is the Yukawa mixing factor, $R$ is the renormalization group running factor, and $\alpha_H$ is the hadronic matrix element. Thus, we obtain a generic estimate for the proton lifetime in GUTs as
\begin{equation}
\tau(p \to e^+ \pi^0) \simeq 7.47 \cdot 10^{35} \left( \frac{M_{\rm GUT}}{10^{16}~\mbox{GeV}} \right)^4 \left( \frac{0.03}{\alpha_{\rm GUT}} \right)^2 \, \mbox{\footnotesize years}.
\end{equation}
Next, let us consider proton decay in basic GUTs and derive an upper bound on the proton lifetime in such GUTs. Assume only dimension-six operators, since other operators can be set to zero in searching for upper bounds. Furthermore, assume that the proton lifetime is induced by superheavy gauge bosons $X$ with mass $M_X$. Using these two assumptions, we obtain an upper bound on the proton lifetime for any GUT with or without SUSY (see {\it e.g.}~Refs.~\cite{Dorsner:2004xa,Nath:2006ut}) as given by
\begin{equation}
\tau_p \lesssim 6.0 \cdot 10^{39} \frac{1}{\alpha_{\rm GUT}^2} \left( \frac{M_X}{10^{16}~\mbox{GeV}} \right)^4 \left( \frac{0.003~\mbox{GeV}^3}{\alpha_{\mbox{\tiny ChPT}}} \right)^2 \, \mbox{\footnotesize years},
\label{eq:taup_estimate}
\end{equation}
where $\alpha_{\mbox{\tiny ChPT}}$ is the hadronic matrix element, stemming from chiral perturbation theory, that is the least-known parameter in the estimate for the upper bound on the proton lifetime in Eq.~\eqref{eq:taup_estimate}. 
In minimal non-SUSY ${\rm SU}(5)$, using two-loop renormalization group running of gauge couplings and Eq.~\eqref{eq:taup_estimate}, the upper bound is found to be \cite{Dorsner:2005ii}
$$
\tau_p \lesssim 1.4 \cdot 10^{36} \, \mbox{years}
$$
with unification at $M_X = 2 \cdot 10^{14}~\mbox{GeV}$ and $\alpha_{\rm GUT} = 1/39$. Thus, non-SUSY GUTs are still allowed with respect to proton decay by the current and best experimental lower bound on the proton lifetime by SK. In fact, in realistic minimal non-SUSY GUTs, it holds that $\tau_p \lesssim 10^{36} \, \mbox{years}$. However, minimal SUSY ${\rm SU}(5)$ is, or is at least very close to be, ruled out.

A historical remark is in place. At the workshop {\it Unified Theories and Baryon Number in the Universe}, KEK (1979) \cite{Watanabe:1979wu}, the following quote was stated in a talk by Y.~Watanabe:
\begin{quote}
{\small Theorists, these days, actually seem to be convinced of the existence of proton decay, but the predicted life time is $\sim 10^{35}$~yr, too stable to be observed by the present-day technology, yet a very challenging problem from experimenta[l]ists' point of view.

Y. Watanabe, {\it Trying to measure the proton's life time}}
\end{quote}
In Tab.~\ref{tab:estimates}, an incomplete list of estimates of predicted proton lifetimes in various models based on different GUTs is presented. 
\begin{table*}
\begin{tabular}{l l c c}
\hline 
\hline
Model class & References & Lifetime [years] & Ruled out?\\
\hline
Minimal ${\rm SU}(5)$ & Georgi \& Glashow \cite{Georgi:1974sy} & $10^{30} - 10^{31}$ & \cellcolor{royes} yes\\
Minimal SUSY ${\rm SU}(5)$ &Dimopoulos \& Georgi \cite{Dimopoulos:1981zb}; Sakai \& Yanagida \cite{Sakai:1981pk} & $10^{28} - 10^{34}$ & \cellcolor{royes} yes\\
SUGRA ${\rm SU}(5)$ & Nath, Chamseddine \& Arnowitt \cite{Nath:1985ub} & $10^{32} - 10^{34}$ & \cellcolor{royes} yes\\
SUSY (MSSM/ESSM) ${\rm SO}(10)/G(224)$ & Babu, Pati \& Wilczek \cite{Babu:1997js} & $2 \cdot 10^{34}$ & \cellcolor{royes} yes\\
SUSY (MSSM/ESSM, $d = 5$) ${\rm SO}(10)$ & Lucas \& Raby \cite{Lucas:1996bc}; Pati \cite{Pati:2003qia} & $10^{32} - 10^{35}$ & \cellcolor{rop} partially\\
SUSY ${\rm SO}(10) + {\rm U}(1)_{\rm fl}$	 & Shafi \& Tavartkiladze \cite{Shafi:1999vm} & $10^{32} - 10^{35}$ & \cellcolor{rop} partially\\
SUSY ($d = 5$) ${\rm SU}(5)$ -- option I & Hebecker \& March-Russell \cite{Hebecker:2002rc} & $10^{34} - 10^{35}$ & \cellcolor{rop} partially\\
SUSY (MSSM, $d = 6$) ${\rm SU}(5)$ or ${\rm SO}(10)$ & Pati \cite{Pati:2003qia} & $\sim 10^{34.9 \pm 1}$ & \cellcolor{rop} partially\\
Minimal non-SUSY ${\rm SU}(5)$ & Dor\v{s}ner \& Fileviez-P{\'e}rez \cite{Dorsner:2005fq} & $10^{31} - 10^{38}$ & \cellcolor{rop} partially\\
Minimal non-SUSY ${\rm SO}(10)$ & & --- & \cellcolor{rono} no\\
SUSY (CMSSM) flipped ${\rm SU}(5)$ & Ellis, Nanopoulos \& Walker \cite{Ellis:2002vk} & $10^{35} - 10^{36}$ & \cellcolor{rono} no\\
GUT-like models from string theory & Klebanov \& Witten \cite{Klebanov:2003my} & $\sim 10^{36}$ & \cellcolor{rono} no\\
Split SUSY ${\rm SU}(5)$ & Arkani-Hamed {\it et al.}~\cite{Arkani-Hamed:2004zhs} & $10^{35} - 10^{37}$ & \cellcolor{rono} no\\
SUSY ($d = 5$) ${\rm SU}(5)$ -- option II & Alciati {\it et al.}~\cite{Alciati:2005ur} & $10^{36} - 10^{39}$ & \cellcolor{rono} no\\
\hline 
\hline
\end{tabular}
\caption{Estimates of predicted proton lifetimes in various GUTs (see Refs.~\cite{Georgi:1974sy,Dimopoulos:1981zb,Sakai:1981pk,Nath:1985ub,Babu:1997js,Lucas:1996bc,Pati:2003qia,Shafi:1999vm,Hebecker:2002rc,Pati:2003qia,Dorsner:2005fq,Ellis:2002vk,Klebanov:2003my,Arkani-Hamed:2004zhs,Alciati:2005ur}). This list is an updated version of the one presented in Ref.~\cite{Bueno:2007um}.}
\label{tab:estimates}
\end{table*}

\section{Proton decay in non-SUSY GUTs}
\label{sec:pdinsg}

In this section, we focus on proton decay in non-SUSY GUTs based on the gauge groups SU(5) and SO(10). First, we discuss an SU(5) model studied in two works, and then, we investigate some SO(10) models.

In Refs.~\cite{Boucenna:2017fna,FileviezPerez:2019ssf}, a minimal non-SUSY SU(5) model has been studied, where the gauge group ${\rm SU}(5) \times {\rm U}(1)_{\rm PQ}$ is broken down to the SM gauge group ${\cal G}_{\rm SM}$ at the GUT scale. First, Boucenna \& Shafi \cite{Boucenna:2017fna} found unification at $M_{\rm GUT} \approx 10^{16} \, \mbox{GeV}$ with the proton lifetime $8 \cdot 10^{34} \, \mbox{years} \lesssim \tau_p \lesssim 3 \cdot 10^{35} \, \mbox{years}$. In this work, axions constitute the dark matter. Then, Fileviez-P{\'e}rez, Murgui \& Plascencia \cite{FileviezPerez:2019ssf} found unification in an interval $M_{\rm GUT} \in (1.12, 10.45) \cdot 10^{15} \, \mbox{GeV}$ that leads to the proton lifetime being in the interval $10^{34} \, \mbox{years} \lesssim \tau_p \lesssim 10^{38} \, \mbox{years}$, which is a larger interval than the one found in Ref.~\cite{Boucenna:2017fna}. Also, in this work, axions constitute the dark matter.

In Ref.~\cite{Lee:1994vp}, four minimal non-SUSY SO(10) models are presented, which have intermediate scales described by gauge groups ${\cal G}$, where the gauge group SO(10) at the unification scale is broken down to the SM gauge group ${\cal G}_{\rm SM}$ in two steps according to ${\rm SO}(10) \longrightarrow {\cal G} \longrightarrow {\cal G}_{\rm SM}$. In Tab.~\ref{tab:ABCD}, the gauge groups ${\cal G}$ of the four models are given as well as the predicted proton lifetimes $\tau_p = \tau(p \to e^+ \pi^0)$ of the different models are listed. Comparing these predicted proton lifetimes with the current and best lower bound from SK, we can conclude that models B and D are allowed and model C is only partially allowed, whereas model A is ruled out. Note that the intermediate gauge groups of models B and D are the Pati--Salam (PS) group ${\rm SU}(4)_C \times {\rm SU}(2)_L \times {\rm SU}(2)_R$ and the left-right (LR) symmetric group ${\rm SU}(3)_C \times {\rm SU}(2)_L \times {\rm SU}(2)_R \times {\rm U}(1)_{B-L}$, respectively.
\begin{table}
\begin{tabular}{l c l}
\hline
\hline
Model & Lifetime [years] & \\
\hline
Model A (${\cal G} = G_{422D}$):	&	$\tau_p = 1.44 \cdot 10^{32.1 \pm 0.7}$ & \cellcolor{royes} ruled out\\
Model B (${\cal G} = G_{422}$):	& 	$\tau_p = 1.44 \cdot 10^{37.4 \pm 0.7}$ & \cellcolor{rono} allowed \\
Model C (${\cal G} = G_{3221D}$):	&	$\tau_p = 1.44 \cdot 10^{34.2 \pm 0.7}$ & \cellcolor{rop} partially \\
Model D (${\cal G} = G_{3221}$):	&	$\tau_p = 1.44 \cdot 10^{37.7 \pm 0.7}$ & \cellcolor{rono} allowed \\
\hline
\hline
\end{tabular}
\caption{Predicted proton lifetimes for four minimal non-SUSY SO(10) models presented in Ref.~\cite{Lee:1994vp}.}
\label{tab:ABCD}
\end{table}

Next, we review a minimal non-SUSY model based on the gauge group ${\rm SO}(10) \times {\rm U}(1)_{\rm PQ}$ \cite{Boucenna:2018wjc}, where ${\rm U}(1)_{\rm PQ}$ is a global Peccei--Quinn (PQ) symmetry. This minimal non-SUSY ${\rm SO}(10)$ model is broken down to the SM gauge group in one step, {\it i.e.}~${\rm SO}(10) \times {\rm U}(1)_{\rm PQ} \longrightarrow {\cal G}_{\rm SM}$. Two color-octet scalar multiplets of $\mbox{\bf 210}_H$ are used in the breaking: $S_1 = (\mbox{\bf 8}, \mbox{\bf 1}, 1)$ and $S_2 = (\mbox{\bf 8}, \mbox{\bf 3}, 0)$. In this model, precise gauge coupling unification can be achieved if $M_1 \leq M_2 \leq M_{\rm GUT}$: $M_1 \simeq 3.10 \cdot 10^3$~GeV, $M_2 \simeq 2.34 \cdot 10^8$~GeV, and $M_{\rm GUT} \simeq 4.51 \cdot 10^{15}$~GeV.
\begin{figure}
\includegraphics[width=0.7\columnwidth]{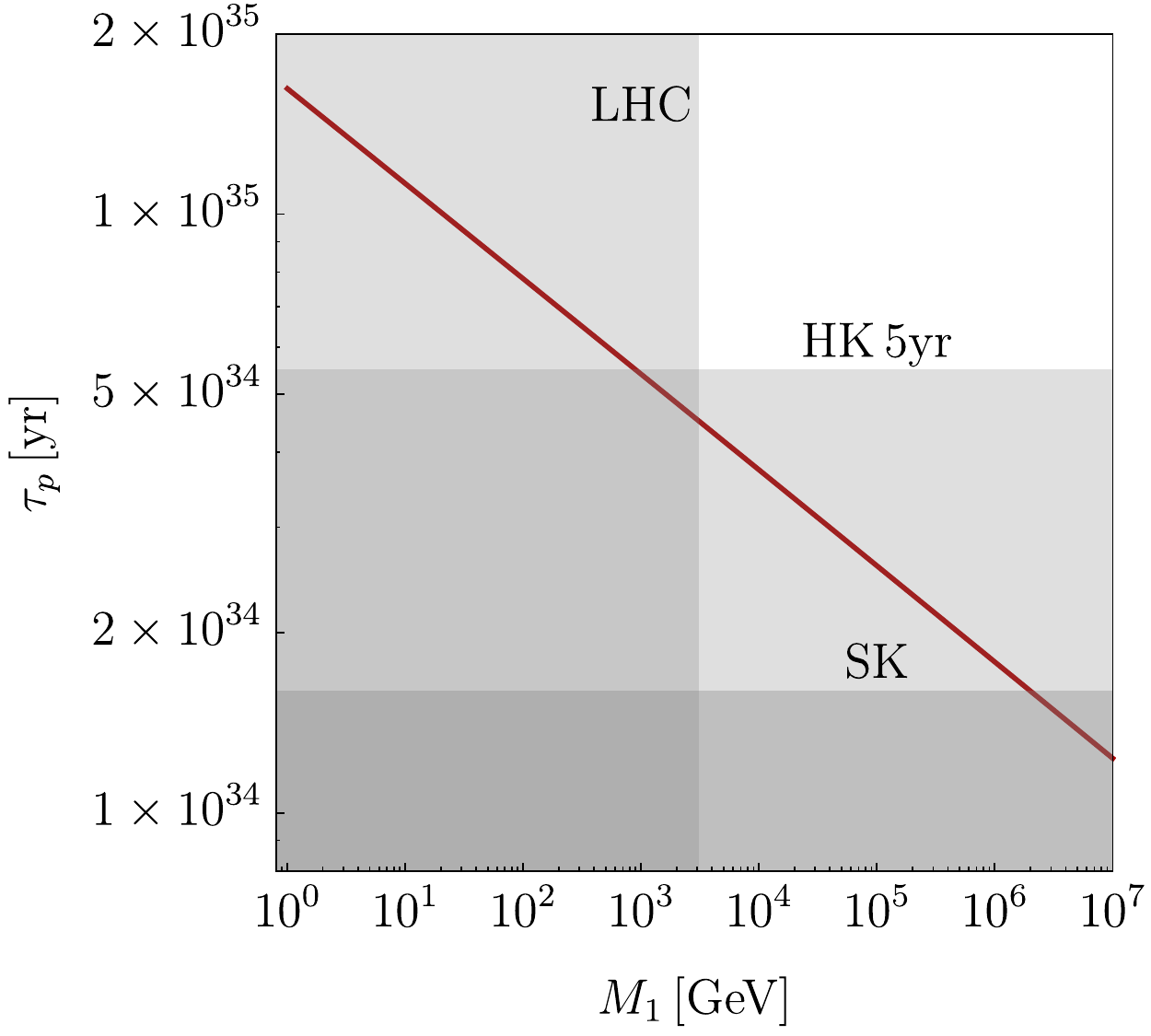}
\caption{Proton lifetime $\tau_p$ as a function of mass $M_1$. This figure has been adopted from Ref.~\cite{Boucenna:2018wjc}. Note that the SK lower bound has increased from $1.6 \cdot 10^{34}$~years to $2.4 \cdot 10^{34}$~years.}
\label{fig:taupM1}
\end{figure}
In Fig.~\ref{fig:taupM1}, the current bounds on $M_1$ from LHC ($M_1 > 3.1 \cdot 10^3$~GeV) and SK ($M_1 < 1.6 \cdot 10^5$~GeV) are displayed, which lead to an allowed proton lifetime in the interval $\tau_p \in (2.4,4.5) \cdot 10^{34}~\mbox{years}$. Note that if the future Hyper-Kamiokande (HK) experiment reaches $\tau_p > 5.5 \cdot 10^{34}$~years in five-years of running, then this model would be strongly disfavored.

Then, we review minimal non-SUSY SO(10) models with one intermediate scale \cite{Meloni:2019jcf}, which are similar to the models studied in Ref.~\cite{Lee:1994vp}. In these minimal non-SUSY SO(10) with one intermediate gauge group (and corresponding scale), seven gauge groups as intermediate gauge groups are analyzed and two out of these groups are found to be allowed by proton decay. In the investigation of these models, renormalization group running is performed at two-loop level and threshold corrections are also taken into account.
\begin{figure}
\includegraphics[height=6cm]{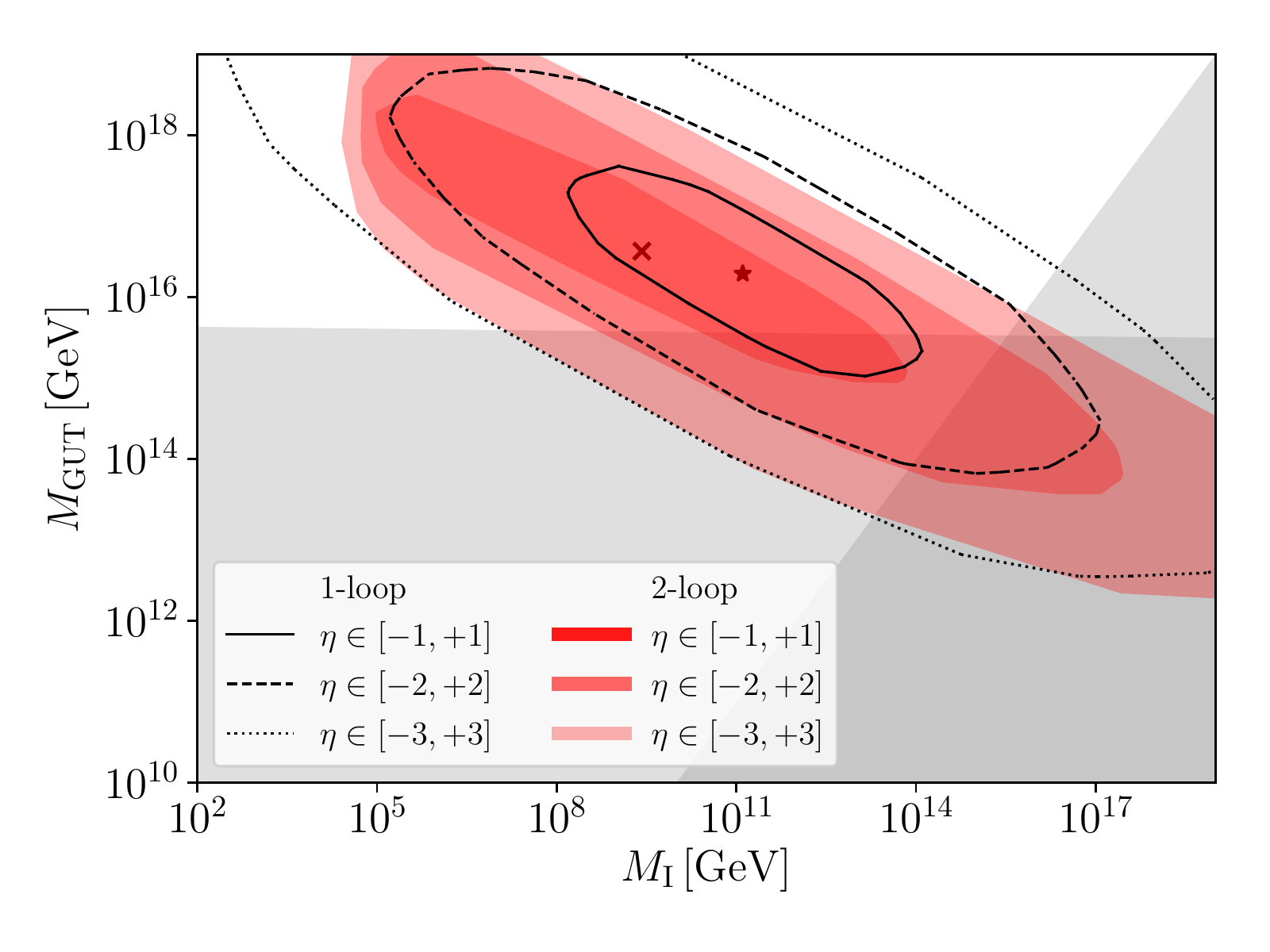}
\includegraphics[height=6cm]{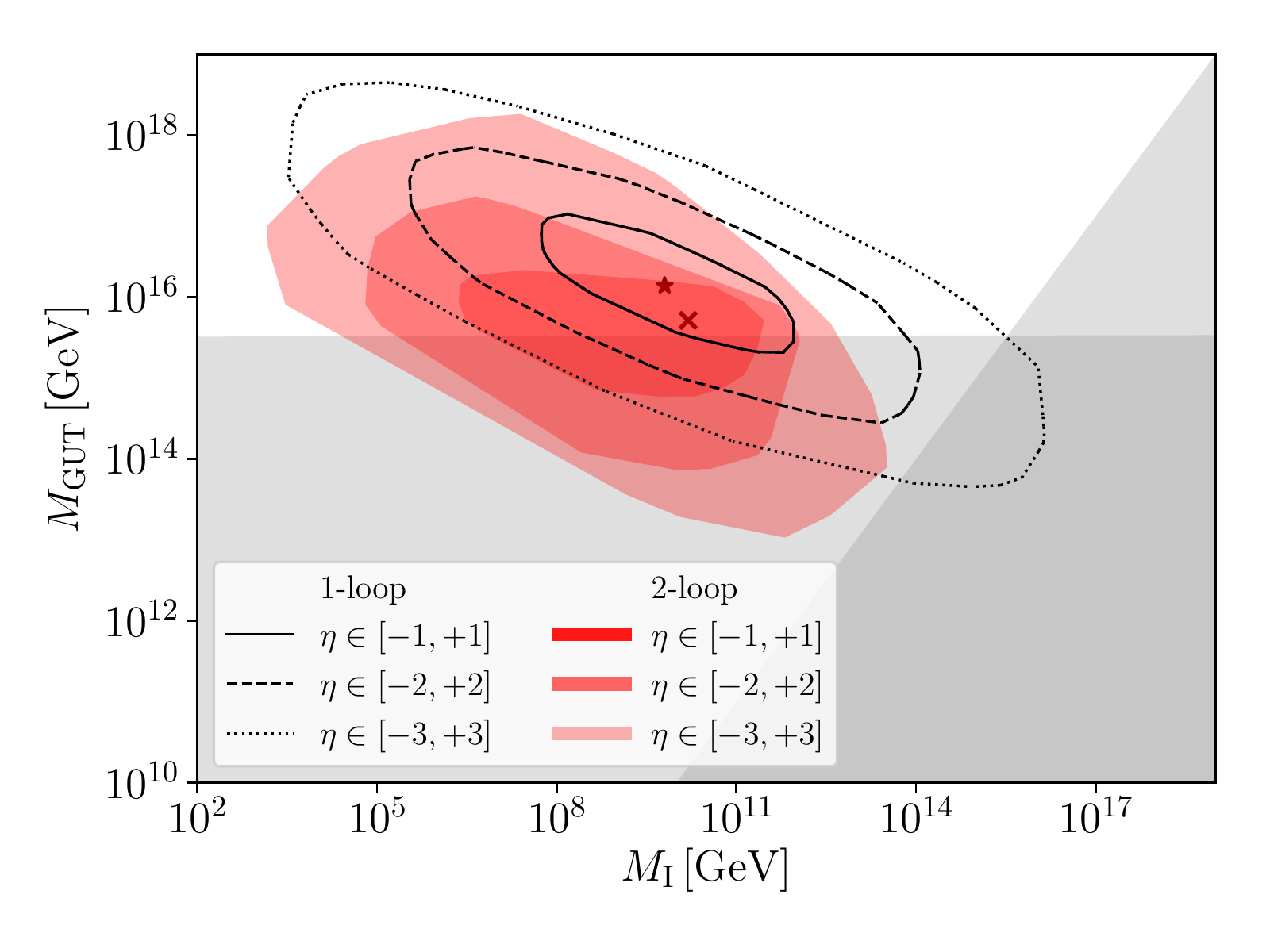}
\caption{Variations in the scales $M_{\rm I}$ and $M_{\rm GUT}$ due to threshold corrections for the Pati--Salam {\it (upper panel)} and left-right symmetric {\it (lower panel)} models which achieve gauge coupling unification. The scales without threshold corrections are marked with a ``$\times$'' (``$\star$'') for two-loop (one-loop) renormalization group running. This figure has been adopted from Ref.~\cite{Meloni:2019jcf}.}
\label{fig:PSLR}
\end{figure}
In Fig.~\ref{fig:PSLR}, we show the variations of the scales for the two allowed models, which are the same as the ones found in Ref.~\cite{Lee:1994vp}, {\it i.e.}~the PS group and the LR symmetric group as intermediate groups, respectively. We find that the model with the PS group as an intermediate group leads to the proton lifetime $\tau_p \simeq 1.2 \cdot 10^{38}~\mbox{years}$, whereas the model with the LR symmetric group as an intermediate group leads to the proton lifetime $\tau_p \simeq 9.4 \cdot 10^{34}~\mbox{years}$.

Finally, we review a minimal non-SUSY SO(10) model with an SU(5) gauge group as intermediate scale \cite{Ohlsson:2020rjc}. This minimal non-SUSY SO(10) model has flipped ${\rm SU}(5) \times {\rm U}(1)_X$ as the intermediate gauge group, which is broken down to the SM gauge group: ${\rm SO}(10) \longrightarrow {\rm SU}(5) \times {\rm U}(1)_X \longrightarrow {\cal G}_{\rm SM}$. This model does not achieve unification if threshold corrections are not taken into account, which are defined as $\eta_i \equiv \ln(M_{S_i}/M_{m \to n})$.
\begin{figure}
\includegraphics[height=6cm]{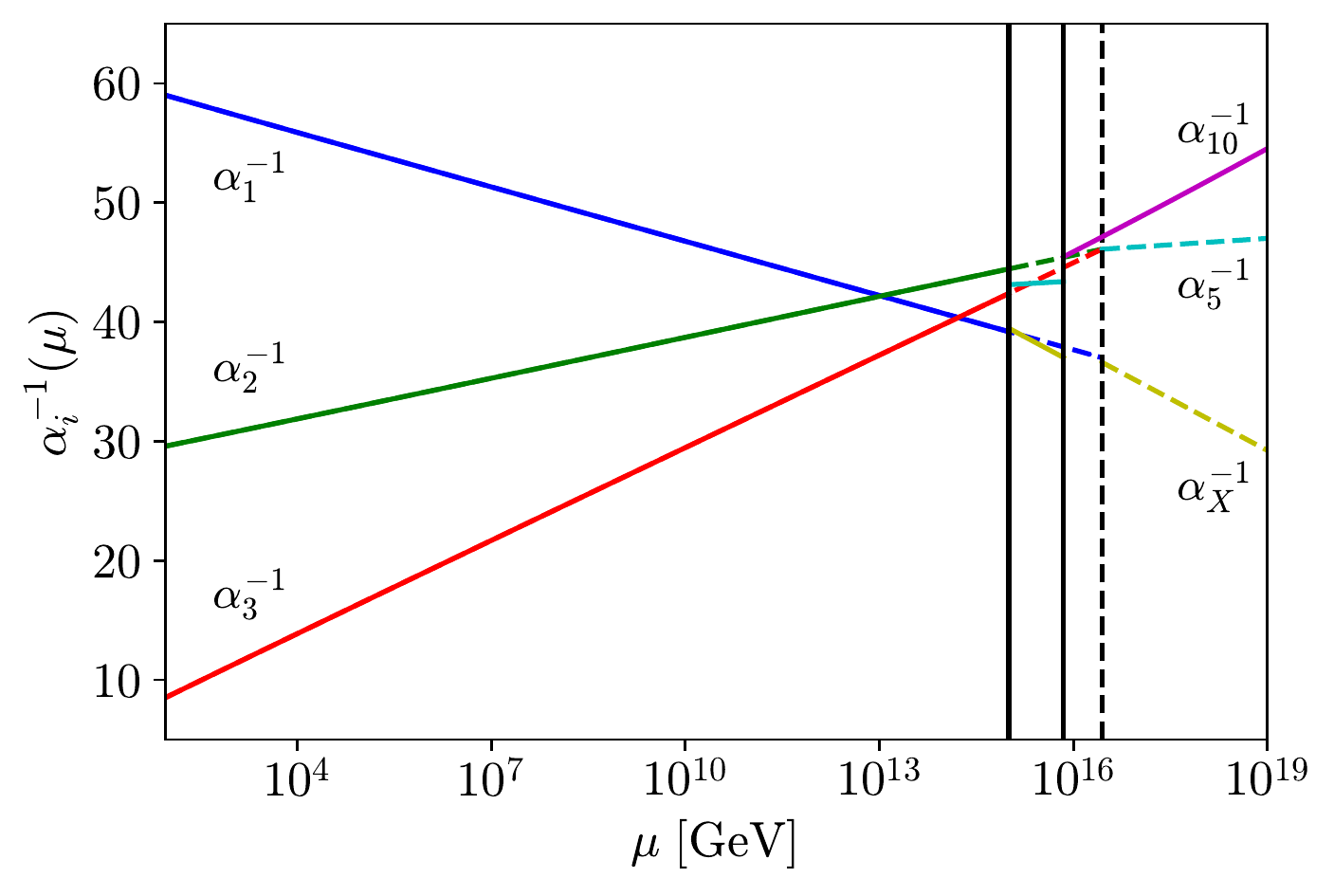}
\includegraphics[height=6cm]{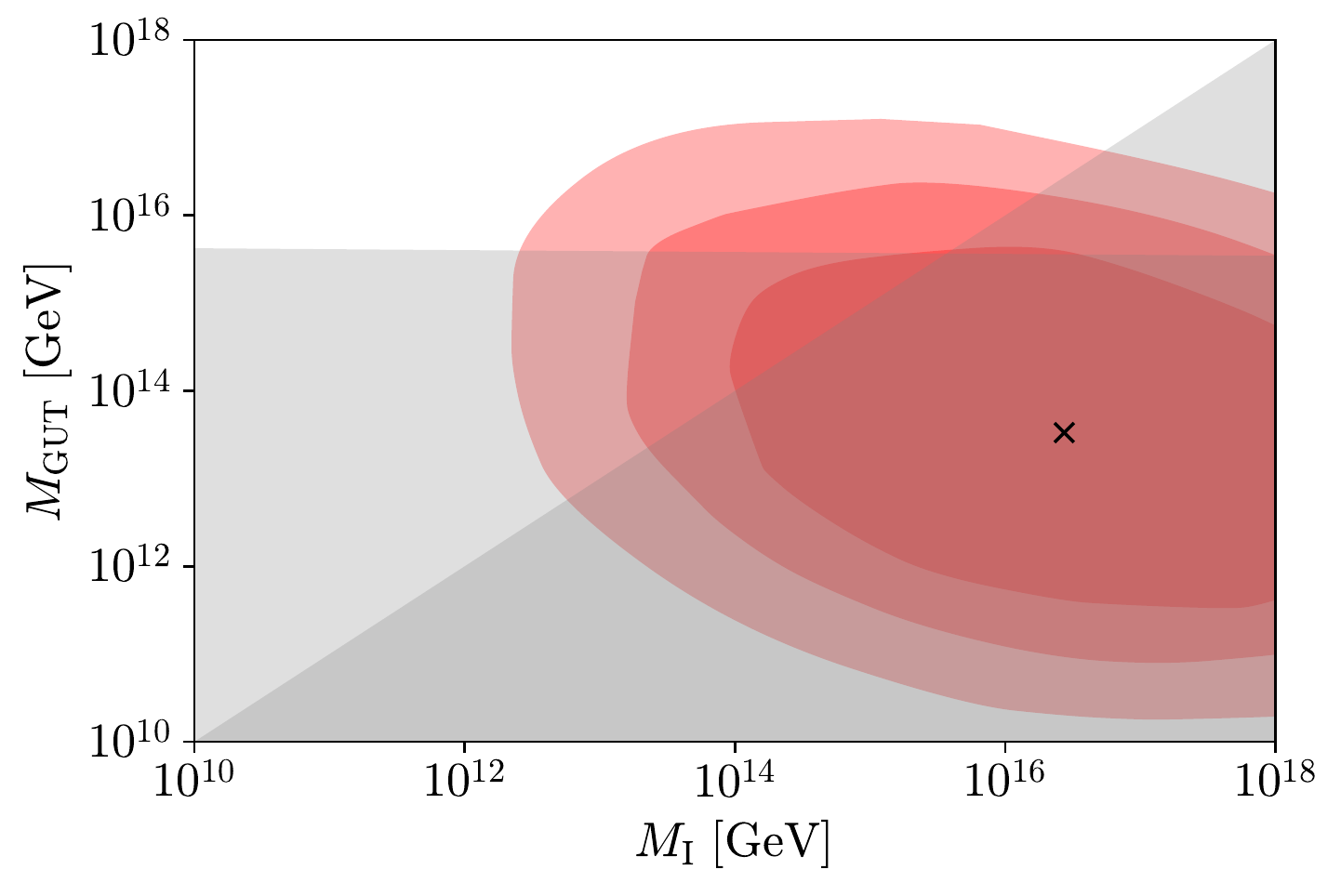}
\caption{{\it Upper panel}: RG running of gauge couplings as a function of the energy scale $\mu$ for the flipped ${\rm SU}(5) \times {\rm U}(1)_X$ model. {\it Lower panel}: Possible scales $M_{\rm I}$ and $M_{\rm GUT}$ with successful unification due to threshold correction in the flipped ${\rm SU}(5) \times {\rm U}(1)_X$ model. These panels have been adopted from Ref.~\cite{Ohlsson:2020rjc}.}
\label{fig:SU(5)xU(1)}
\end{figure}
In the upper panel of Fig.~\ref{fig:SU(5)xU(1)}, we display a possible unification of the gauge couplings in this model, which can {\it e.g.}~be achieved by the choices of parameter values: $M_{\rm I} \simeq 10^{15}$~GeV and $M_{\rm GUT} \simeq 7 \cdot 10^{15}$~GeV. In addition, in the lower panel of Fig.~\ref{fig:SU(5)xU(1)}, we show the threshold corrections that are needed in order to achieve unification and allow proton decay. In fact, we observe that $\eta_i \in [-4,4]$ (``middle'' red color) are needed, which are indeed large threshold corrections.

\section{In the last five years: 2017--now}
\label{sec:2017now}

In this section, the development of proton decay in GUTs based on SU(5) and SO(10) in the last five years in the literature is summarized and presented. This presentation does not included the models that have already been discussed in Sec.~\ref{sec:pdinsg}.

First, in SU(5) GUTs, the development of proton decay includes the following works: 
\begin{itemize}
\item Lee \& Mohapatra \cite{Lee:2016wiy}: non-SUSY ${\rm SU}(5) \times {\rm SU}(5)$, $\tau_{p \to e^+ \pi^0} \simeq 6.2 \cdot 10^{34} \, \mbox{years}$
\item Fornal \& Grinstein \cite{Fornal:2017xcj}: ${\rm SU}(5)$, stable proton
\item Rehman, Shafi \& Zubair \cite{Rehman:2018nsn}: SUSY flipped ${\rm SU}(5)$, $\tau_p \sim 10^{36} \, \mbox{years}$
\item Haba, Mimura \& Yamada \cite{Haba:2018vvu}: non-SUSY ${\rm SU}(5)$
\item Ellis {\it et al.}~\cite{Ellis:2020qad}: SUSY (flipped) ${\rm SU}(5)$
\item Mehmood, Rehman \& Shafi \cite{Mehmood:2020irm}: SUSY flipped ${\rm SU}(5) \times {\rm U}(1)_R$
\item Babu, Gogoladze \& Un \cite{Babu:2020ncc}: minimal SUSY ${\rm SU}(5)$, $\tau_{p \to \overline{\nu} K^+} \lesssim 10^{35} \, \mbox{years}$
\item Dor\v{s}ner, D\v{z}aferovi{\'c}-Ma\v{s}i{\'c} \& Saad \cite{Dorsner:2021qwg}: realistic minimal non-SUSY ${\rm SU}(5)$
\item Evans \& Yanagida \cite{Evans:2021hyx}: minimal SUSY (CMSSM, $d = 5$) ${\rm SU}(5)$
\item Haba \& Yamada \cite{Haba:2021rzs}: SUSY ($d = 5$) flipped ${\rm SU}(5)$
\item Ellis {\it et al.} \cite{Ellis:2021vpp}: SUSY flipped ${\rm SU}(5)$,  $\tau_p \gtrsim 10^{36} \, \mbox{years}$
\end{itemize}
Thus, considering this list of SU(5) models, SUSY flipped SU(5) seems to be popular.

Second, in SO(10) GUTs, the development of proton decay includes the following works: 
\begin{itemize}
\item Babu, Bajc \& Saad \cite{Babu:2016bmy}: minimal non-SUSY ${\rm SO}(10) \to \mbox{PS}$
\item Mohapatra \& Severson \cite{Mohapatra:2018biy}: SUSY ${\rm SO}(10)$
\item Babu, Fukuyama, Khan \& Saad \cite{Babu:2018qca}: SUSY ${\rm SO}(10) \times {\rm U}(1)_{\rm PQ}$, $\tau_{p \to e^+ \pi^0} \simeq 9 \cdot 10^{34} \, \mbox{years}$
\item Haba, Mimura \& Yamada \cite{Haba:2019wwt}: SUSY ${\rm SO}(10)$
\item Chakraborty, Parida \& Sahoo \cite{Chakraborty:2019uxk}: minimal non-SUSY ${\rm SO}(10)$
\item Hamada {\it et al.}~\cite{Hamada:2020isl}: non-SUSY ${\rm SO}(10) \to \mbox{LR}$
\item King, Pascoli, Turner \& Zhou \cite{King:2021gmj}: non-SUSY ${\rm SO}(10)$ with all possible intermediate scales
\item Preda, Senjanovi\'{c} \& Zantedeschi \cite{Preda:2022izo}: minimal non-SUSY ${\rm SO}(10)$, $\tau_p \lesssim 10^{35} \, \mbox{years}$
\end{itemize}
Thus, considering this list of SO(10) models, non-SUSY SO(10) seems to be popular.

\bigskip

\vfill

\section{Future experiments}
\label{sec:fe}

The four most promising future experiments to search for proton decay are:
\begin{itemize}
\item {\sc JUNO} (Jiangmen, China; under construction, data taking in 2023): 20~kton liquid scintillator detector.\\
{\sc JUNO} has a possibility to search for proton decay.
\item {\sc Hyper-Kamiokande} (Gifu, Japan; under construction, data taking in 2027): 188~kton water-Cherenkov detector ($\sim 8~\times$~SK).\\
To search for proton decay is among the main objectives for {\sc Hyper-Kamiokande}.
\item {\sc DUNE} (Illinois \& South Dakota, USA): 68~kton liquid Argon detector.\\
{\sc DUNE} has a possibility to search for proton decay.
\item {\sc ESSnuSB} (Sweden): 0.5~Mton water-Cherenkov detector ($\sim 20~\times$~SK).\\
{\sc ESSnuSB} has an excellent opportunity to search for proton decay.
\end{itemize}

\section{Summary and outlook}
\label{sec:s&o}

In summary, there are in general many different available GUTs: minimal/non-minimal, SUSY/non-SUSY, SU(5)/SO(10), {\it etc.} In particular, I have not been able to review them all. Personally, I believe that GUTs without SUSY seem to be more viable than those with SUSY. Especially, minimal non-SUSY SO(10) GUTs are allowed and seem prosperous. It should be mentioned that any reasonable and testable model must be able to survive the present experimental limits on proton decay. In conclusion, future experiments with huge detectors will increase the lower bound on proton lifetime, but may eventually also detect proton decay.

\begin{acknowledgments}
I thank the organizers of Neutrino 2022 for inviting me to give the talk on which this work is based.

T.O.~acknowledges support by the Swedish Research Council (Vetenskapsr{\aa}det) through Contract No.~2017-03934.
\end{acknowledgments}

\vfill

\bibliographystyle{apsrev4-2}
\bibliography{neutrino2022_proton_decay3}

\end{document}